\documentclass[journal]{IEEEtran}

\usepackage{amssymb, graphicx, cite, amsmath, psfrag, minibox, mathrsfs, bigints, stfloats, color, upgreek, gensymb, epstopdf, textcomp, amsthm, amsfonts, euscript, calligra}

\usepackage[font=scriptsize]{subcaption}
\usepackage[font=scriptsize]{caption}

\usepackage{mathtools, amsfonts, multirow, array}
\usepackage{verbatim, mdwtab, relsize, cleveref, mdwmath}
\usepackage{algorithmicx}
\usepackage{algpseudocode}
\usepackage[ruled]{algorithm2e}

\usepackage{setspace}
\usepackage{float}

\usepackage{tikz}
\usetikzlibrary{shapes,arrows}
\usepackage{textcomp}

\def\pioh{\widehat{\pi}_0}
\def\pilh{\widehat{\pi}_1}
\def\P0{P^{(0)}}

\def\Pbar{\overline{P}_{\text{av}}}
\def\Ibar{\overline{I}_{\text{av}}}

\def\SUTx{SU$_{\text{tx}}$}
\def\SURx{SU$_{\text{rx}}$}

\def\SNR0{$\text{SNR}^{(0)}$}
\def\SNR1{$\text{SNR}^{(1)}$}

\ifCLASSINFOpdf
\else
\fi

\begin{document}
\title{On the Spectrum Sensing, Beam Selection and Power Allocation in Cognitive Radio Networks Using Reconfigurable Antennas}

\author{Hassan Yazdani,
        Azadeh~Vosoughi,~\IEEEmembership{Senior Member,~IEEE} \\
        University of Central Florida\\
        E-mail: {\tt \normalsize h.yazdani@knights.ucf.edu,  azadeh@ucf.edu} 
}


%
%



\markboth{}%
{Shell \MakeLowercase{\textit{et al.}}: Bare Demo of IEEEtran.cls for Journals}
%

\newcounter{MYeqcounter}
\multlinegap 0.0pt                     



\maketitle

 

\begin{abstract}

In this paper, we consider a cognitive radio (CR) system consisting of a primary user (PU) and a pair of secondary user transmitter (\SUTx) and secondary user receiver (\SURx).  The \SUTx ~is equipped with a reconfigurable antenna (RA) which divides the angular space into $M$ sectors. The RA chooses one sector among $M$ sectors for its data transmission to \SURx. The \SUTx ~first senses the channel and monitors the activity of PU for a duration of $T_\text{sen}$ seconds. We refer to this period as {\it{channel sensing phase}}.  Depending on the outcome of this phase, \SUTx ~stays in this phase or enters the next phase, which we refer to as {\it{transmission phase}}. The transmission phase itself consists of two phases: {\it{channel training phase}} followed by {\it{data transmission phase}}. During the former phase, \SUTx ~sends pilot symbols to enable channel training and estimation at \SURx. The \SURx ~selects the best beam (sector) for data transmission and feeds back the index of the selected beam as well as its corresponding channel gain. We also derive the probability of determining the true beam and take into account this probability in our system design. During the latter phase, \SUTx ~sends data symbols to \SURx ~over the selected beam with constant power $\Phi$ if the gain corresponding to the selected beam is bigger than the threshold $\zeta$. We find the optimal channel sensing duration $T_\text{sen}$, the optimal power level $\Phi$ and a optimal threshold $\zeta$, such that the ergodic capacity of CR system is maximized, subject to average interference and power constraints. In addition, we derive closed form expressions for outage and symbol error probabilities of our CR system.
\end{abstract}
%
%

\IEEEpeerreviewmaketitle

\section{Introduction}\label{Se1}

The explosive rise in demand for high data rate wireless applications has turned the spectrum into a scarce resource. Cognitive radio (CR) is a promising solution which alleviates spectrum scarcity problem by allowing an unlicensed or secondary user (SU) to access licensed bands in a such way that its imposed interference on license holder primary users (PUs) is limited  \cite{Arslan, Nalla, Liang2, Gursoy1, Beaulieu}.  

With the help of directional antennas, a system designer can go beyond just filling the spectrum holes in time and/or frequency domains and can further improve the spectral efficiency (well beyond what is attainable with omni-directional antennas), via utilizing the spectrum white spaces in spatial (angular) domain and allowing a PU and a SU use simultaneously the same frequency band, by steering their directional antenna beams to non-interfering directions. Consider an opportunistic CR network where SUs are equipped with directional antennas. SUs can identify the spectrum holes in spatial (angular) domain and use the identified unused directions for data communication. This significantly enhances the spectrum utilization, compared to the scenario where SUs in the same network are equipped with omini-directional antennas \cite{ICASSPpaper, AsilomarPaper, GlobalSIP2}. 

Reconfigurable antennas (RAs), with the capabilities of dynamically modifying their characteristics (e.g., operating frequency, radiation pattern, polarization) are emerging as promising solutions to efficiently utilize the spatial (angular) domain and beam steering in wireless communication systems, including CR networks \cite{Mahmoud}. RA has been used for identifying the directional spectrum sensing opportunities in CR networks as well as for directional wireless and millimeter wave communication systems and surveillance \cite{Rohollah, Jafarkhani}. 
An electrically steerable parasitic array radiator (ESPAR) antenna is a special kind of RAs, that has been used for identifying the spectral holes in spatial domain in CR networks. ESPAR divides the angular domain into several sectors (beams) and switches between beampatterns of sectors in a time-division fashion (only one of $M$ beams is active at a time). For CR networks, the RAs can provide an improved spectrum sensing and transmit/receive capability, due to a signal-to-noise ratio (SNR) increase for transmission and reception of directional signals, and can limit out-of-band interference to and from PUs \cite{Spatial_SS_Parasitic}. It has been demonstrated that RAs have the ability to transmit multiple data streams by projecting on beamspace basis \cite{Random_Aerial}. Also, they can be used for blind interference alignment through beampattern switching \cite{BlindInterference}. RAs have been used for performance enhancement of multiple-input multiple-output (MIMO) systems, via exploiting the additional degree of freedom provided by beam selection and enabling joint beam and antenna selection optimization \cite{Ghrayeb2, Ghrayeb1, Evans}.

In this paper, we consider a CR system consisting of a PU and a pair of SU transmitter (\SUTx) and SU receiver (\SURx).  The \SUTx ~is equipped with a reconfigurable antenna with the capability of choosing one sector among $M$ sectors for its data transmission to \SURx. The \SUTx ~first senses the channel and monitors the activity of PU for a duration of $T_\text{sen}$ seconds. We refer to this period as {\it{channel sensing phase}}.  Depending on the outcome of this phase, \SUTx ~stays in this phase or enters the next phase, which we refer to as  {\it{transmission phase}}. The transmission phase itself consists of two phases: {\it{channel training phase}} followed by {\it{data transmission phase}}. During the former phase, \SUTx ~sends pilot symbols to enable channel training and estimation at \SURx. The \SURx ~selects the best beam (sector) for data transmission and feeds back the index of the selected beam as well as its corresponding channel gain. We also derive the probability of determining the true beam and take into account this probability in our system design. During the latter phase, \SUTx ~sends data symbols to \SURx ~over the selected beam with constant power $\Phi$ if the gain corresponding to the selected beam is bigger than a threshold $\zeta$. Our objective is to find the optimal channel sensing duration $T_\text{sen}$, the optimal power level $\Phi$ and the optimal threshold $\zeta$, such that the ergodic capacity of CR system is maximized, subject to average interference and power constraints. In addition, we derive closed form expressions for outage and symbol error probabilities of our CR system.
%
%
%
\section{System Model}\label{Se2}

\par
\begin{figure}[!t]
\centering
\psfrag{g}[Bl][Bl][0.6]{$g$}
\psfrag{gss}[Bl][Bl][0.6]{$g_\text{ss}$}
\psfrag{gsp}[Bl][Bl][0.6]{$g_\text{sp}$}
\psfrag{SUtx}[Bl][Bl][0.7]{\SUTx}
\psfrag{SUrx}[Bl][Bl][0.7]{\SURx}
\psfrag{PU}[Bl][Bl][0.7]{PU}
\psfrag{phiSR}[Bl][Bl][0.6]{$\phi_\text{SR}$}
\psfrag{phiPU}[Bl][Bl][0.6]{$\phi_\text{PU}$}
\includegraphics[width=40mm]{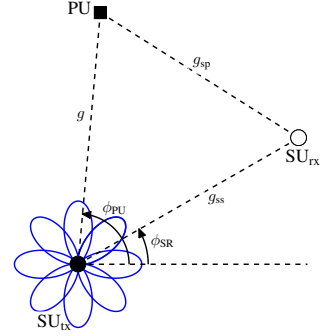}
\caption{Our cognitive radio system with reconfigurable antenna.} 
\label{SystemModelFig}      
\vspace{-0mm}  
\end{figure}

\begin{figure}[!t]
\centering
\psfrag{k1}[Bl][Bl][0.7]{$\kappa_1$}
\psfrag{k2}[Bl][Bl][0.7]{$\kappa_2$}
\psfrag{k3}[Bl][Bl][0.7]{$\kappa_3$}
\psfrag{k4}[Bl][Bl][0.7]{$\kappa_4$}
\psfrag{k5}[Bl][Bl][0.7]{$\kappa_5$}
\psfrag{k6}[Bl][Bl][0.7]{$\kappa_6$}
\psfrag{k7}[Bl][Bl][0.7]{$\kappa_7$}
\psfrag{k8}[Bl][Bl][0.7]{$\kappa_8$}
\includegraphics[width=40mm]{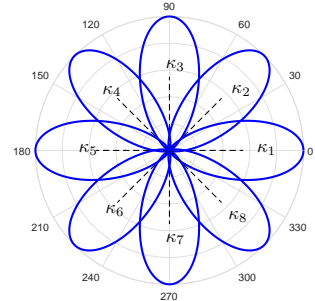}
\caption{Beampatterns  of a reconfigurable antenna with $8$ sectors.} 
\label{Pattern_fig}      
\end{figure}

 Our CR system model is shown in Fig. \ref{SystemModelFig}, consisting of a PU and a pair of \SUTx ~and \SURx. We note that PU in our system model can be a primary transmitter or receiver. Similar to \cite{Varshney}, we assume when PU is active it is engaged in a bidirectional communication with another PU. The  other PU is located far from \SUTx ~and its activity is not considered in this paper. The \SUTx ~is equipped with a RA (for both channel sensing and communication) with the capability of choosing one sector among $M$ sectors for its data transmission to \SURx, while \SUTx ~and PU use omni-directional antennas \footnote{Throughout this paper, ``sector'' and ``beam'' are used interchangeably.}. Similar to \cite{Cabric}, we model the radiation pattern of every sector of \SUTx's antenna in $x\!-\!y$ (azimuth) plane, with the Gaussian pattern as
%
\begin{equation}\label{pattern_formula}
p(\phi) =  A_1 + A_0 e^{ -B \left ( \frac{ \mathcal{M(\phi)}} {\phi_{\text{3dB}}} \right )^2 },
\end{equation}
%
where
%
\begin{equation}
\mathcal{M(\phi)} = \text{mod}_{2 \pi} ( \phi + \pi ) - \pi,
\end{equation}
%
$\text{mod}_{2 \pi} ( \phi  )$ denotes the remainder of $\frac{\phi}{2 \pi}$,  $B=\ln(2)$, $\phi_{3\text{dB}}$ is the half-power beam-width, {$A_1$} and $A_0$ are two constant antenna parameters. We set $A_1=L A_0$ where $L \ll 1$ is the antenna loss in side lobe. We denote the radiation pattern of $m$-th sector in angle $\phi$ by 
%
\begin{equation}
p_m(\phi ) = p(\phi - \kappa_m)
\end{equation}
%
where $\kappa_m = \frac{2 \pi (m-1)}{M}$. In Fig. \ref{Pattern_fig}, the beampatterns of a RA with $8$ sectors are shown. The orientation of PU and \SURx ~with respect to \SUTx ~are denoted by $\phi_\text{PU}$  and $\phi_\text{SR}$, receptively, and we assume that \SUTx ~knows $\phi_\text{SR}$. The fading coefficients from PU to \SUTx,  \SUTx ~to \SURx ~and PU to \SURx ~are denoted by $h$, $h_\text{ss}$ and $h_\text{sp}$, respectively, if \SUTx ~uses omni-directional antenna. We model the fading coefficients as circularly symmetric complex Gaussian random variables, and,  $g=|h|^2$, $g_\text{ss}=|h_\text{ss}|^2$ and $g_\text{sp}=|h_\text{sp}|^2$ are mutually independent  exponentially distributed random variables  with mean $\gamma$, $\gamma_\text{ss}$ and $\gamma_\text{sp}$, respectively. 
%
%
%
%
In our problem, we assume that SUs and PU cannot cooperate and SUs cannot estimate $g$ and $g_\text{sp}$. However, we assume that \SUTx ~knows mean values $\gamma$ and $\gamma_\text{sp}$. Let $\psi_m$ and $\chi_m$ denote the fading coefficients of channel between $m$-th sector of \SUTx ~and PU, and between $m$-th sector of \SUTx ~and \SURx, respectively, where  $\psi_m = h  \sqrt{p_m(\phi_\text{PU} )} $, $\chi_m= h_{\text{ss}} \sqrt{p_m(\phi_\text{SR} )}$ and $p_m(\phi_\text{PU})$ and $p_m(\phi_\text{SR})$ indicate the radiation pattern of $m$-th sector at angles $\phi_\text{PU}$ and $\phi_\text{SR}$, respectively. Also, we assume that the channel gain $\nu_m = |\chi_m|^2$ is an exponential random variable with mean $\delta_m$, and \SUTx ~knows $\delta_m$, for all $m$ \cite{ArslanESPAR}. 
%
\section{Our Problem Statement}\label{}

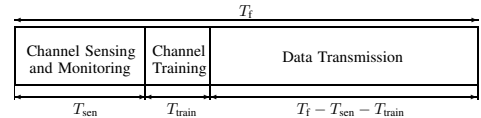
\begin{figure}[!t]
\vspace{0mm}
\centering
\hspace{-0mm}
 \vspace*{0pt}
\setlength{\unitlength}{3.2mm} 
\centering
\scalebox{0.60}{
\begin{picture}(27,9)
\hspace{0mm}
\centering
\put(-2.55,-0.5){\line(0,1){5}}
\put(-2.5,.5){\framebox(32,4){}}
\put(-1.7,2.5){Channel Sensing}
\put(-1.5,1.3){and Monitoring}
\put(6.5,-0.5){\line(0,1){5}}
\put(7.0,2.5){Channel}
\put(7.0,1.3){Training}
\put(11.0,-0.5){\line(0,1){5}}
\put(16,2.1){Data Transmission}
\put(29.52,-0.5){\line(0,1){5}}

\put(-2.55,5.0){\vector(1,0){32.1}}
\put(-2.55,5){\vector(-1,0){0}}
\put(13,5.3){$T_\text{f}$}

\put(-2.55,-0.35){\vector(1,0){9.1}}
\put(-2.55,-0.35){\vector(-1,0){0}}
\put(1.5,-1.5){$T_\text{sen}$}

\put(6.5,-0.35){\vector(1,0){4.5}}
\put(6.5,-0.35){\vector(-1,0){0}}
\put(8,-1.5){$T_\text{train}$}

\put(11.0,-0.35){\vector(1,0){18.5}}
\put(11.0,-0.35){\vector(-1,0){0}}
\put(17,-1.5){$T_\text{f} - T_\text{sen} - T_\text{train}$}

\end{picture}}
\vspace{5mm} 
\caption{The structure of frame employed by \SUTx.} 
\label{FrameStructure}
\end{figure}

We suppose the SUs employ a frame with a fixed duration of $T_\text{f}$ seconds, depicted in Fig. \ref{FrameStructure}. We assume \SUTx ~first senses the channel and monitors the activity of PU. We refer to this period as {\it{channel sensing phase}} (with a variable duration of $T_\text{sen}$ seconds). $T_\text{sen}$ is the sensing time duration for all $M$ sectors, i.e., every sector senses the channel for $T_\text{sen}/M$ seconds. Depending on the outcome of this phase, \SUTx ~stays in this phase or enters the next phase, which we refer to as  {\it{transmission phase}}. The transmission phase itself consists of two phases: {\it{channel training phase}} (with a fixed duration of $T_\text{train}$  seconds) followed by {\it{data transmission phase}} (with a variable duration of $T_\text{f} -\! T_\text{sen} -\! T_\text{train}$ seconds). During the former phase, \SUTx ~sends pilot symbols to enable channel training and estimation at \SURx. During the latter phase, \SUTx ~sends data symbols to \SURx. Given $T_\text{f}$ and $T_\text{train}$  we have $0< T_\text{sen} < T_\text{f} -T_\text{train}$. In the following, we describe how \SUTx ~operates during these three distinct phases. Based on these descriptions,  we provide our problem statement.
%
%
%
%
%
%
\subsection{Channel Sensing Using RA}\label{GLRT_SS}

During this phase, \SUTx ~senses the channel and monitors the activity of PU. Suppose $\mathcal{H}_1$  and $\mathcal{H}_0$  represent the binary hypotheses of PU being active and inactive, respectively, with prior probabilities $\Pr \{\mathcal{H}_1\}= \pi_1$ and $\Pr \{\mathcal{H}_0\}=\pi_0$. \SUTx ~applies a binary detection rule to decide whether or not PU is active. Let $\widehat{\mathcal{H}}_1$ and $\widehat{\mathcal{H}}_0$ denote the detector outcome. 
When active, PU transmits signal $s(t)$ with power $P_p$. We assume \SUTx ~collects $N= \lfloor  T_\text{sen}/(M T_\text{s}) \rfloor$ samples at each sector, where $T_\mathrm{s}$ is the sampling period. We denote the discrete-time symbol received in $m$-th sector of \SUTx ~at time instant $t=nT_\mathrm{s}$ as
%
\begin{align*}
y_m(n) = & \psi_m (n)  s(n) + w_m(n),
\end{align*}
%
We model the  transmitted signal $s(n)$ by PU as $s(n) \sim \mathcal{CN}(0,P_p)$. The term $w_m(n)$ is the additive noise at $m$-th sector of \SUTx's antenna and is modeled as $w_m(n) \sim \mathcal{CN}(0,\sigma_w^2) $. It is assumed that $\psi_m(n), w_m(n)$ and $s(n)$ are mutually independent. We assume $w_m(n)$ are independent and thus uncorrelated in both time and  space domains, i.e., $\mathbb{E}\left \{ {w}_m(i) {w}^*_{m^\prime}(i^\prime) \right \} = \sigma_w^2 \delta[m \! - \! m^\prime] \delta[i \! - \! i^\prime]$ where $\delta[\cdot]$ is Kronecker's delta. Since only one beam is active at a time, the channel gains $\psi_m(n)$ are uncorrelated in both time and space domains, i.e., $\mathbb{E} \left \{ {\psi}(n) {\psi}^*(n^\prime)\right \} =  E_A \gamma \delta[m\!-\!m^\prime] \delta[n\!-\!n^\prime]$, where $E_A \!=\! \frac{1}{2\pi}\int_{0}^{2\pi} \!\! p(\theta) d\theta$. 
The hypothesis testing problem at $n$-th time instant for $m$-th sector is then given by
%
\begin{equation*} 
\begin{cases}
 {\cal H}_{0}: &  y_m(n) =  w_m(n), \\
 {\cal H}_{1}: & y_m(n) = \psi_m(n)  s(n) + w_m(n). 
\end{cases}
\end{equation*}
Suppose \SUTx ~uses an energy detector to detect the activity of PU and let $\varepsilon_m$ be the energy of received signal at sector $m$. We have
%
\begin{equation}
\varepsilon_m = \frac{1}{N} \sum_{n=1}^{N} |y_m(n)|^2.
\end{equation}
%
We consider the summation of energies of received signals over all sectors as the decision statistics as:
%
\begin{align}
T = \frac{1}{M} \sum_{m=1}^M \varepsilon_m 
\gtreqless
\begin{matrix}
\widehat{\mathcal{H}}_1 \cr \widehat{\mathcal{H}}_0
\end{matrix}
\eta.
\end{align}
%
%
Under hypothesis $\mathcal{H}_0$, for large $N$ we invoke the central limit theorem (CLT), to approximate $T$ as Gaussian with distribution $T \sim  \mathcal{N}  ( \sigma_w^2, \sigma^2_{T|{\cal H}_0} )$, where
%
\begin{equation*}
\sigma^2_{T|{\cal H}_0} = \frac{\sigma_w^4}{MN}.
\end{equation*}
%
Similarly, under hypothesis $\mathcal{H}_1$ for large $N$, $T$ can be approximated with another Gaussian with the distribution $T \sim  \mathcal{N}  (\mu, \sigma^2_{T|{\cal H}_1})$ where $\mu = P_p \gamma E_A + \sigma_w^2$ and
%
\begin{align*}\label{sigma2_H1}
\sigma^2_{T|{\cal H}_1} = & \frac{1}{MN} \Big [ \sigma_w^4 + 2 \gamma P_p E_A \sigma_w^2 +\gamma^2 P_p^2 \big (3 E_B  -  MN E_A^2 \big ) \Big ] \\  & +  \frac{\gamma^2 P_p^2}{M^2} \sum_{m=1}^{M} \sum_{m^\prime=1}^{M} E_{m m^\prime},
\end{align*}
%
and $E_{m m^\prime}\! = \!\frac{1}{2 \pi} \int_{0}^{2\pi} \! p_m(\theta) p_{m^\prime}(\theta) d\theta$ and $E_B = E_{mm}$.
Then, the false alarm and detection probabilities of this energy detector are given as follows
%
\begin{equation*}
\begin{array}{ll}
P_\text{fa} = Q \left(\frac{\eta - \sigma_w^2}{\sigma_{T|{\cal H}_{0}}} \right ),  ~~~~& P_\text{d} = Q \left( \frac{\eta -\mu}{ \sigma_{T|{\cal H}_{1}}} \right ).
\end{array}%
\end{equation*}
%
For a given value of $P_\text{d} = \overline{P}_\text{d}$, the probability of false alarm can be written as
%
\begin{equation}\label{}
P_\text{fa} = Q \left( \frac{  \sigma_{T|{\cal H}_1} Q^{-1} \big (\overline{P}_\text{d} \big ) + \mu -\sigma^2_w }{\sigma_{T|{\cal H}_0 }} \right ).
\end{equation}
%
%
where $Q(\cdot)$ is the Q-function. Therefore, the probabilities of events $\widehat{ \mathcal{H}}_0$ and $\widehat {\mathcal{H}}_1$ become $\pioh = \Pr \{ \widehat{ \mathcal{H}}_0 \} = \pi_1 (1 \! -\! \overline{P}_\text{d} ) + \pi_0 (1 \! - \! {P}_\text{fa}) $ and $\pilh = \Pr \{ \widehat{ \mathcal{H}}_1 \} = \pi_1 \overline{P}_\text{d}  + \pi_0 {P}_\text{fa} $, respectively.
The accuracy of channel sensing impacts the maximum information rate that \SUTx ~can transmit reliably to \SURx. Our problem formulation incorporates the effect of imperfect channel sensing on the constrained ergodic capacity maximization. 
%
%
%
\subsection{Estimating the orientation of PU ($\phi_\text{PU}$)}

As long as the channel is sensed busy, \SUTx ~stays in {\it{channel sensing phase}}. While being in this phase, \SUTx ~estimates  the location (orientation) of  PU using one of the methods explained in \cite{Directional_SS_ESPAR, MILCOM, Jabal, Cabric, Cabric2}.
\SUTx ~uses $\phi_\text{PU}$ for adapting its transmit power during {\it{data transmission phase}}. We note that, there is a non-zero error when \SUTx ~estimates $\phi_\text{PU}$ and we can incorporate this error in our system design and performance analysis. However, in this paper we assume that \SUTx ~can estimate $\phi_\text{PU}$ perfectly (with no error).
%
%
%
\subsection{Determining the Sector Corresponding to \SURx ~and Finding the Strongest Channel between \SUTx-\SURx }\label{Se3}

\begin{figure}[!t]
\vspace{-0mm}
\centering
	\begin{subfigure}[b]{0.25\textwidth}                
		\centering	              
		\psfrag{g}[Bl][Bl][0.6]{$g$}
		\psfrag{gss}[Bl][Bl][0.6]{$g_\text{ss}$}
		\psfrag{gsp}[Bl][Bl][0.6]{$g_\text{sp}$}
		\psfrag{SUtx}[Bl][Bl][0.7]{\SUTx}
		\psfrag{SUrx}[Bl][Bl][0.7]{\SURx}
		\psfrag{PU}[Bl][Bl][0.7]{PU}
		\psfrag{phiSR}[Bl][Bl][0.6]{$\phi_\text{SR}$}
		\psfrag{phiPU}[Bl][Bl][0.6]{$\phi_\text{PU}$}
		\includegraphics[width=37mm]{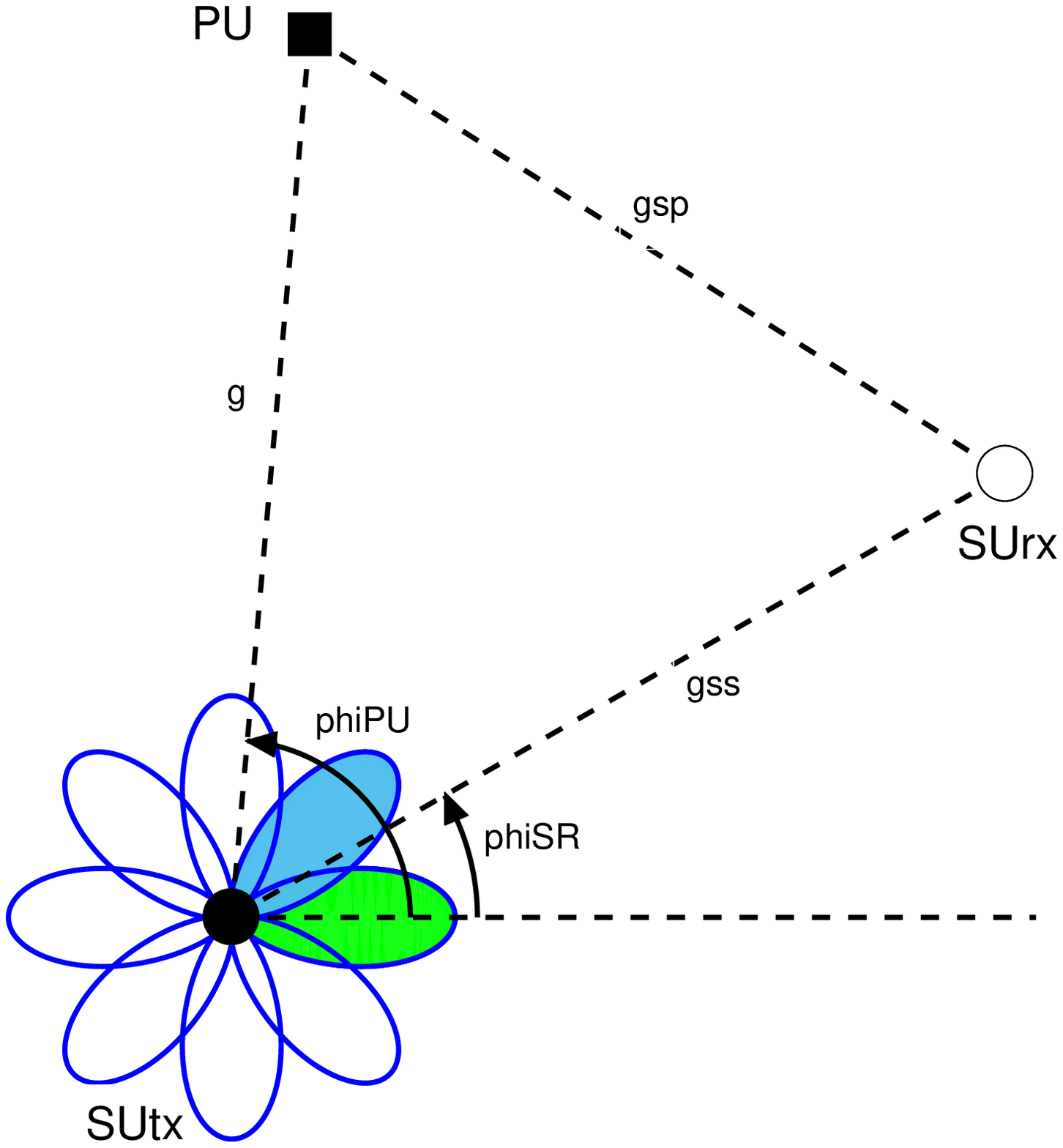}
		\caption{} 
		\label{i_SR}      	       
	\end{subfigure}%
      \begin{subfigure}[b]{0.25\textwidth}
 		\centering      
		\psfrag{g}[Bl][Bl][0.6]{$g$}
		\psfrag{gss}[Bl][Bl][0.6]{$g_\text{ss}$}
		\psfrag{gsp}[Bl][Bl][0.6]{$g_\text{sp}$}
		\psfrag{SUtx}[Bl][Bl][0.7]{\SUTx}
		\psfrag{SUrx}[Bl][Bl][0.7]{\SURx}
		\psfrag{PU}[Bl][Bl][0.7]{PU}
		\psfrag{phiSR}[Bl][Bl][0.6]{$\phi_\text{SR}$}
		\psfrag{phiPU}[Bl][Bl][0.6]{$\phi_\text{PU}$}
		\includegraphics[width=37mm]{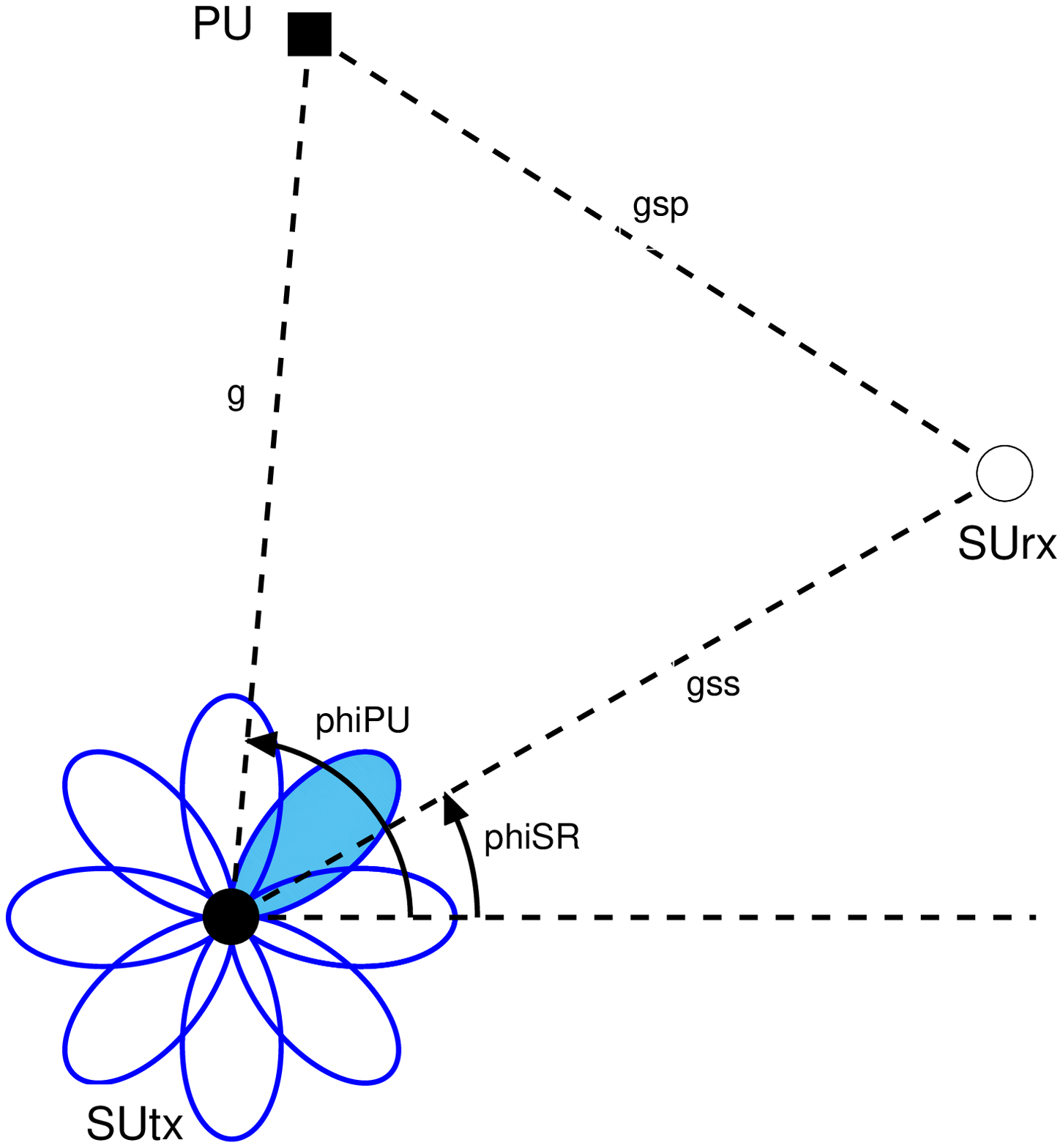}
		\caption{} 
		\label{i_SR_Sel}   
      \end{subfigure} \\
\caption{A schematic to show how \SUTx ~selects the strongest channel between \SUTx ~and \SURx ~(a) \SURx ~is located between the first and second beams (sectors), (b) The second sector is selected for data transmission ($m_\text{SR}^*=2$).}
\vspace{0mm}
\end{figure}

When the channel is sensed idle, \SUTx ~leaves {\it{channel sensing phase}} and enters {\it{channel training phase}}. Suppose \SUTx ~knows \SURx ~is located between two adjacent sectors, however, it does not know which one is better, in terms of enabling a higher transmission rate. During channel training phase, \SUTx ~sends pilot symbols over these two sectors, to enable channel training and estimation at \SURx. Without loss of generality, suppose \SURx ~is located between the first and second sectors as shown in Fig. \ref{i_SR}. Using the  received training signal, \SURx ~estimates the channel gains  $\nu_m = |\chi_m|^2, m=1, 2$ and determines the strongest channel $\nu^*=\max \{\nu_1, \nu_2 \}$ and the corresponding beam index $m_\text{SR}^* = {{ \arg\max}} ~\{ \nu_1, \nu_2 \}$. For example in Fig. \ref{i_SR_Sel}, we have $m_\text{SR}^*=2$, i.e., the second beam has the strongest channel gain. Then, \SURx ~feeds back $m_\text{SR}^*$  as well as the value of $\nu^*$ to \SUTx. We take into account the probability of determining the true sector corresponding to \SURx ~on the constrained capacity maximization.
\par We denote the correlation coefficient between channel gains of sectors $m$ and $m^\prime$ by $\rho_{mm^\prime}$. The correlation coefficient depends on the structure and design of RA \cite{Ghrayeb1} and we assume \SUTx ~has complete knowledge of the correlation coefficient $\rho_{mm^\prime}$.
%
Let $\rho$ represent  the correlation coefficient between two adjacent sectors which \SURx ~is located. The joint probability distribution function (PDF) and cumulative density function (CDF) of channel gains $\nu_1$ and $\nu_2$ can be written as following
%
\begin{equation}\label{}
f_{\nu_1 \nu_2}(y_1 , y_2)  = \frac{\alpha_1}{ \delta_2}   e^{ -\alpha_1 y_1 }  e^{ -\alpha_2 y_2 }   I_0 \Big ( 2\sqrt{ \rho^2 \alpha_1 \alpha_2 y_1 y_2}  \Big)
\end{equation}
%
\begin{align}
F_{\nu_1 \nu_2}(y_1 , y_2)   = & 1- e^{ -\frac{y_1}{\delta_1} } Q_1 \Big(\sqrt{2 \alpha_2 y_2},\sqrt{2 \rho^2 \alpha_1 y_1} \Big) \nonumber \\
 -  & e^{ -\frac{y_2}{ \delta_2} } \left [ 1- Q_1 \Big(\sqrt{2 \rho^2 \alpha_2 y_2},\sqrt{2 \alpha_1 y_1} \Big)    \right ]
\end{align}
%
where $\alpha_m = \frac{1}{ \delta_m (1-\rho^2)}$ for $m=1, 2$, $I_0(\cdot)$ is the modified Bessel function of the first kind and $Q_1(\cdot, \cdot)$ is the Marcum Q-function. Then the CDF of selected gain $\nu^*$ is $F_{\nu^*}(x) = F_{\nu_1 \nu_2}(x,x)$ and the PDF is 
%
\begin{align}\label{f_v}
f_{\nu^*} (x) =  \frac{1}{ \delta_1} e^{-\frac{x}{ \delta_1}}    \left [ 1- Q_1 \Big(\sqrt{2 \rho^2 \alpha_1 x},\sqrt{2 \alpha_2 x} \Big)  \right ] \nonumber \\
+  \frac{1}{ \delta_2} e^{-\frac{x}{ \delta_2}}    \left [ 1- Q_1 \Big(\sqrt{2 \rho^2 \alpha_2 x},\sqrt{2 \alpha_1 x} \Big)  \right ].
\end{align}
%
The probability of the first beam being selected when \SURx ~is between the first  and  second  beams becomes
%
\begin{align}\label{Prob_i}
\Delta_1 = & \Pr \big \{ m^*_\text{SR} =1 \big \} = \Pr \big (\nu_1 > \nu_{2} \big ) \nonumber \\
= & \! \int_{0}^{\infty} \!\! \int_{0}^{x} f_{\nu_1, \nu_2} (x, y) dy dx \!=\! \frac{1}{\alpha_2 \delta_2} \sum_{k=0}^{\infty} \rho^{2k} \bigg [ 1\! -\! \left (\!  \frac{\alpha_1}{\alpha_2} \! \right )^{k+1} \nonumber \\  
& \times \frac{\Gamma(2k\!+\!2)}{(k\!+\!1) (k!)^2}    ~{}_2 F_1 \Big (k\!+\!1, 2(k\!+\!1); k\!+\!2 ~; \frac{-\alpha_1}{\alpha_2} \Big ) \bigg ]
\end{align}
%
where ${}_2 F_1 (\cdot, \cdot; \cdot; \cdot)$ is the hypergeometric function \cite{BookTable}. In special case when $\rho=0$, we have  $\Delta_1 = \frac{\delta_1}{\delta_1 + \delta_2}$.
%
%
%
%
%
\subsection{Data Transmission}

When the channel is sensed idle, \SUTx ~sends data to \SURx ~over the selected sector ($m_\text{SR}^*$) using the following power control policy
%
\begin{equation}\label{Power1}
P(\nu^*) = \left\{
\begin{array}{ll} 
\Phi,  & \text{if}~~ \nu^* \geq \zeta   \\
0, & \text{if}~~ \nu^* < \zeta \\
\end{array}\right.  
\end{equation}
%
According to \eqref{Power1}, when the selected channel (sector) is too weak ($\nu^*$ is less than the cut-off threshold $\zeta$) \SUTx ~does not transmit data. However, when channel is strong enough, \SUTx ~sends data with constant power $\Phi$. We find $\Phi$ and $\zeta$ such that the ergoic capacity of \SUTx-\SURx ~link is maximized. The ergodic capacity of \SUTx-\SURx ~link is \cite{GlobalSIP2}
%
\begin{equation}\label{C_Ergodic}
C= D_t \mathbb{E} \big \{ \alpha_{0} c_{0,0} + \beta_0 c_{1,0}  \big \},
\end{equation}
%
where $D_t = (T_\text{f}-T_\text{sen}-T_\text{train})/T_\text{f}$ is the fraction of time in which \SUTx ~sends data to \SURx, $c_{i,0}$ is the instantaneous capacity of this link corresponding to the event $\mathcal{H}_i$ and $\widehat{\mathcal{H}}_0$, given as
%
\begin{align}
c_{0,0} = & \log_2 \left(1+\frac{ \nu^* P(\nu^*)}{\sigma^2_{w}}\right) \label{c0i}, \\ 
c_{1,0} = & \log_2 \left(1+\frac{ \nu^* P(\nu^*)}{\sigma^2_{w}+P_{p}  g_\text{sp} {\color{red} } }\right), \label{c1i}
\end{align}
%
%
and $\alpha_0=\pi_0(1-P_\text{fa})$ ,  $\beta_0 = \pi_1 (1-\overline{P}_\text{d})$. Let $\Ibar$ indicate the maximum allowed interference power imposed on PU and $\Pbar$ denote the maximum allowed average transmit power of \SUTx. To satisfy the average interference constraint (AIC), we have 
%
\begin{equation}\label{Iav0}
D_t \beta_0\mathbb{E}  \big \{ g_\text{sp}  ~p( \kappa_{\text{SR} }^* \! - \! \phi_{\text{PU}}) P(\nu^*) \big \} \leq \Ibar,
\end{equation}
%
and to satisfy the average power constraint (APC), we have
%
\begin{equation}\label{Pav0}
 D_t \pioh \mathbb{E} \big\{  P(\nu^*) \big\} \leq \Pbar. 
\end{equation}
%
\par Notice that for a given $T_\text{f}$, if we increase the sensing time $T_\text{sen}$, the spectrum sensing will be more accurate  and data communication between \SUTx ~and \SURx ~would cause less interference on PU. On the other hand, the available time for data transmission to \SURx ~decreases. Therefore, a trade-off exists between sensing and transmission capacity in terms of $T_\text{sen}$. Our main objective is to find the optimal channel sensing duration $T_\text{sen}$, the optimal power level $\Phi$ and the optimal threshold $\zeta$, such that the ergodic capacity $C$ in \eqref{C_Ergodic} is maximized, subject to AIC and APC given in \eqref{Iav0} and \eqref{Pav0}, respectively. In other words, we are interested in solving the following constrained optimization problem
%
\begin{align}\tag{P1}\label{Prob1}
{\underset { T_\text{sen}, \Phi, \zeta}  {\text{Maximize}} }  ~C = D_t \mathbb{E} \big \{ \alpha_{0} C_{0,0} + \beta_0 C_{1,0}  \big \}& \nonumber
\end{align}
\begin{align*}
\begin{array}{ll}
\text{s.t.:}  ~~& 0  < T_\text{sen} < T_\text{f}  \!-\!T_\text{train},\nonumber \\
 &   ~\Phi \geq 0, \zeta \geq 0, \nonumber \\
 & \eqref{Iav0}  ~\text{and} ~ \eqref{Pav0} ~\text{are satisfied.} \nonumber
 \end{array}
\end{align*}
%
%
%
\section{Formalizing and Solving \eqref{Prob1}}

Since SUs and PU cannot cooperate, \SUTx ~cannot estimate the channel gain $g_\text{sp}$ and thus $c_{1,0}$ cannot be directly maximized at \SUTx. Instead, we consider a lower bound on its average over $g_\text{sp}$, denoted as $\mathbb{E}_{g_\text{sp}}  \{c_{1,0} \}$.
Using the Jensen's inequality \cite{Cover}, the lower bound on $\mathbb{E}_{g_\text{sp}} \! \left\{c_{1,0} \right \}$ becomes
%
\begin{equation}
\mathbb{E}_{g_\text{sp}}  \left\{c_{1,0} \right \} \geq  \log_2 \left(1+\frac{ \nu^* P(\nu^*)}{\sigma^2_{w}+ \sigma_\text{p}^2 }\right) = c_{1,0}^\text{LB}
\end{equation}
%
where $\sigma^2_{p}  =  P_{p}  \gamma_\text{sp}$. Let  $C^\text{LB}= D_t \mathbb{E}_{\nu^*} \! \big \{ \alpha_{0} c_{0,0} + \beta_0 c_{1,0}^\text{LB}  \big \}$ where $C^\text{LB}$ is the lower bound on $C$ in \eqref{C_Ergodic}. From now on, we focus on $C^\text{LB}$. 
Next, we focus on the AIC in \eqref{Iav0} and find the term $\mathbb{E} \{ p( \kappa_{\text{SR}}^* \! - \! \phi_{\text{PU}} )  \}$. By using the average probabilities derived  in  \eqref{Prob_i} we have 
%
\begin{equation*}
\mathbb{E} \big\{  p \big (\kappa_{i_\text{SR}^*} - {\phi}_\text{PU} \big ) \big\}  = {\Delta}_1 ~p \big ( \kappa_1 \! - \! {\phi}_\text{PU}  \big )  + {\Delta}_2 ~p \big ( \kappa_2 \!- \! {\phi}_\text{PU}  \big ).
\end{equation*}
%
%
By defining $b_0  = \beta_0 \gamma_\text{sp} \big [ {\Delta}_1 p \big ( \kappa_1 \! - \! {\phi}_\text{PU}  \big )  + {\Delta}_2 p \big ( \kappa_2 \!- \! {\phi}_\text{PU}  \big ) \big ]$,
%
%
%
the optimal power $\Phi$, given $T_\text{sen}$ and $\zeta$ can be easily obtained as
%
\begin{equation}\label{PHI}
\Phi =   \frac{1 }{ \Big(1-F_{\nu^*} (\zeta) \Big)} \min \left \{  \frac{\Pbar}{D_t \pioh} , \frac{\Ibar}{D_t b_0} \right\}.
\end{equation}
%
To obtain a closed form for ergodic capacity, we expand the PDF in \eqref{f_v} using the following equation
%
\begin{equation*}
Q_1(\sqrt{\lambda},\sqrt{y}) =  \sum_{j=0}^{\infty} \frac{  {(\frac{\lambda}{2})}^j  }{j !} e^{-\frac{(y+\lambda)}{2}} \sum_{i=0}^{j} \frac{  (\frac{y}{2})^i }{i !}.
\end{equation*}
%
and rewrite $f_{\nu^*}(x)$ as the following form
%
\begin{equation}
f_{\nu^*}(x) = \sum_{m=1}^{2} \frac{1}{ \delta_m} e^{-\frac{x}{\delta_m}} - \sum_{j=0}^{\infty}  \sum_{i=0}^{j} D_{ij} x^{i+j}  e^{-x \omega}
\end{equation}
%
where $\omega = \alpha_1+\alpha_2$ and $D_{ij} = \frac{ \rho^{2 j}} {j ! ~i !}  \Big ( \frac{\alpha_1^j \alpha_2^i}{ \delta_1} + \frac{\alpha_1^i \alpha_2^j}{ \delta_2}  \Big )$.
%
To solve \eqref{Prob1}, we consider two initial values for $\zeta$ and $T_\text{sen}$ and obtain $\Phi$ using \eqref{PHI}. 
Then, $\zeta^\text{opt}$ and $T_\text{sen}^\text{opt}$ can be obtained by maximizing the ergodic capacity given in \eqref{CLB} using searching methods such as bisection, where
%
\begin{figure*}
\begin{align}\label{CLB}
C^\text{LB}_\text{Opt} =  \underset{\zeta , T_\text{sen}} {\max}  ~~\frac{ D_t}{\ln(2)}\!  \Bigg [  \sum_{m=1}^{2} \Big [ \alpha_0 G \big (\delta_m, \text{SNR}^{(0)} ,\zeta \big ) + \beta_0 G \big (\delta_m, \text{SNR}^{(1)} ,\zeta) \Big ]  - \sum_{j=0}^{\infty} \sum_{i=0}^{j}  D_{ij} \Big [\alpha_0 V \big (i\!+\!j, \omega , \text{SNR}^{(0)} ,\zeta \big )  \nonumber \\
 + \beta_0 V \big (i\!+\!j, \omega , \text{SNR}^{(1)} ,\zeta \big )  \Big]   \Bigg] 
\end{align}
\hrulefill
\end{figure*}
%
%
\begin{align*}
\begin{array}{ll}
\text{SNR}^{(0)} =  \frac{\Phi }{\sigma_{w}^2}, ~~~~~~~& \text{SNR}^{(1)} =  \frac{\Phi }{\sigma_{w}^2+\sigma_{p}^2},
\end{array}
\end{align*}
%
\begin{align*}
G \big (\delta,\text{S},\zeta \big ) = & e^{-\frac{\zeta}{\delta} } \ln \Big (1\!+\!\text{S} \zeta \Big )-e^{\frac{1}{\delta \text{S}} } \text{Ei} \Big( \frac{- 1}{ \delta \text{S} }  \big (1\!+\! \text{S} \zeta  \big) \Big ),
\end{align*}
%
%
and $\text{Ei}(\cdot)$ is the exponential integral \cite{BookTable} and $V(\cdot)$ can be calculated using the recursive equation in \eqref{V_func}. In \eqref{H_func}, $\Gamma(\cdot,\cdot)$ is the incomplete Gamma function. 
%
%
\begin{figure*}
\begin{subequations}\label{V_func}
%
\begin{equation}
V \big (n,\omega,\text{S},\zeta \big) =  \frac{n}{\omega} V \big (n\!-\!1,\omega,\text{S},\zeta \big) + \frac{\zeta^n}{\omega}  G \big (1/\omega,\text{S},\zeta \big )  + \frac{n}{\omega} e^{\frac{\omega}{\text{S}}} H \big (n\!-\!1,\omega,\text{S},\zeta \big) 
\end{equation}
%
\begin{equation}\label{H_func}
H \big (k,\omega,\text{S},\zeta \big) =  \sum_{j=0}^{k} {k \choose j} \frac{ (-\text{S})^{j-k} }{ (j\!+\!1) ~\omega^{j+1}} \left [ \Big (\omega \zeta \! +\! \frac{\omega}{S} \Big )^{j+1} ~\text{Ei} \Big (\!\! -\! \omega \zeta \! - \! \frac{\omega} {\text{S}} \Big) + \Gamma \Big (j\!+\!1, \omega \zeta\!+\! \frac{\omega} {\text{S}} \Big )  \right ] 
\end{equation}
\end{subequations}
\hrulefill
\end{figure*}
%
%
%
%
%
\section{Outage and Symbol Error Probabilities }
Two other relevant metrics to evaluate the performance of our CR system with the RA at \SUTx ~are outage probability and symbol error probability (SEP), denoted as $P_\text{out}$ and $P_\text{e}$, respectively. We define $P_\text{out}$ as the probability of \SUTx ~not transmitting data due to the weak \SUTx-\SURx  ~channel. In the following, we derive closed-form expressions for $P_\text{out}$ and $P_\text{e}$. The outage probability $P_\text{out}$ can be directly obtained using the CDF of $\nu^*$ as
%
\begin{equation}
P_\text{out} = \Pr \big \{ P(\nu^*) \!= \! 0 \big \}  = F_{\nu^*}(\zeta).
\end{equation}
%
For many digital modulation schemes SEP can be written as a function with the following form  \cite{Evans}
%
\begin{equation}\label{Pe00}
P_\text{e} = \mathbb{E}  \left \{ Q(\sqrt{\Psi ~\text{SNR}_\text{rx}}) \right \},
\end{equation}
%
\noindent where $\Psi$  is a constant parameter related to the type of modulation and $\text{SNR}_\text{rx}$ is the received signal-to-noise-ratio (SNR) at \SURx ~given as
%
\begin{equation}\label{}
\text{SNR}_\text{rx} = \left\{
\begin{array}{lll} 
\frac{\Phi \nu^* }{\sigma_w^2},  & \text{if}~~ \nu^* \geq \zeta, ~(\mathcal{H}_0,\widehat{\mathcal{H}}_0)  \\
\frac{\Phi \nu^* }{\sigma_w^2+\sigma_p^2},  & \text{if}~~ \nu^* \geq \zeta, ~(\mathcal{H}_1,\widehat{\mathcal{H}}_0)   \\
0, & \text{else} \\
\end{array}\right.  
\end{equation}
%
After some manipulation, SEP can be written as \eqref{SEP1}, where 
%
%
\begin{equation*}
J \big(k,\text{S},\Psi, \omega ,\zeta \big) \!=\! \frac{1}{\text{S}^{k}} \Big ( \frac{\omega}{\text{S} \Psi}\!+\!\frac{1}{2} \Big )^{-k-\frac{1}{2}} ~\Gamma \Big (k\!+\!\frac{1}{2} , \big (\frac{\text{S} \Psi}{2} \! +\! \omega \big )\zeta \Big )
\end{equation*}
%
and $E_{ij} = \frac{\alpha_1^i \alpha_2^j}{i! ~j !}  \left ( \rho^{2j} - \rho^{2 i}  \right ) $.
%
%
%
\begin{figure*}
\begin{multline}\label{SEP1}
P_{\text{e}}= \frac{1}{2 \sqrt{2\pi}} \Bigg [ \bigg [ \alpha_0 J \Big (0, \text{SNR}^{(0)} ,\Psi,0,\zeta \Big )   + \beta_0 J \Big (0, \text{SNR}^{(1)} ,\Psi,0,\zeta \Big ) \bigg ] \Big (1-F_{\nu^*}(\zeta) \Big)  - \alpha_0  J \Big (0, \text{SNR}^{(0)} ,\Psi,\frac{1}{ \delta_2 },\zeta \Big )   \\
 - \beta_0  J \Big (0, \text{SNR}^{(1)} ,\Psi,\frac{1}{ \delta_2 },\zeta \Big )    \sum_{j=0}^{\infty} \sum_{i=0}^{j} \frac{E_{ij}}{\Psi^{i+j}} \bigg ( \alpha_0 J \Big (i\!+\!j, \text{SNR}^{(0)} ,\Psi, \omega,\zeta \Big )  +   \beta_0  J \Big (i\!+\! j, \text{SNR}^{(1)} ,\Psi, \omega,\zeta \Big )  \bigg )   \Bigg ]
\end{multline}
\hrulefill
\end{figure*}
%
%
%
\section{Simulation Results}

In this section, we illustrate the effect of reconfigurable antennas on the ergodic capacity and outage and symbol error probabilities of our secondary network  by Matlab simulations. Assume $\sigma^2_w\!=\!1$, $\gamma_\text{ss}\!= \!\gamma_\text{sp}\!=\!\gamma\! = \!1$, $\pi_1\! =\! 0.4$, $L=0.01$, $T_\text{f}=10$ ms, $f_s = 100$ KHz, 
$\overline{P}_\text{d}=0.85$, $P_p \!=\!0.2$ watts, $\Psi=4$.  

\par
\begin{figure}[!t]
\vspace{-0mm}
\centering
	\begin{subfigure}[b]{0.245\textwidth}                
		\centering	    		
		\psfrag{phi  =  25}[Bl][Bl][0.34]{$\phi_\text{3dB}\!=\!25^\degree$}
		\psfrag{phi  =  35}[Bl][Bl][0.34]{$\phi_\text{3dB}\!=\!35^\degree$}		
		\psfrag{Omni}[Bl][Bl][0.34]{Omni}	
		\includegraphics[width=34mm]{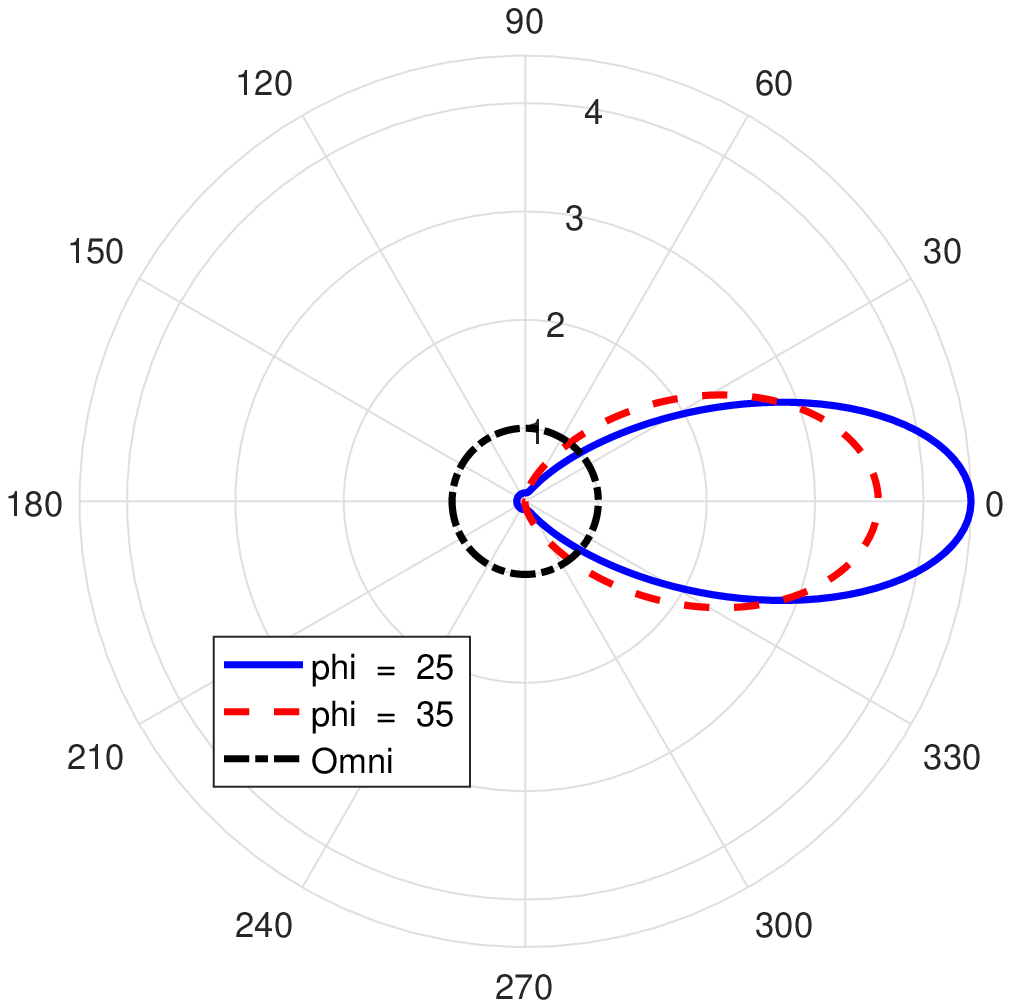}
		\caption{} 
		\label{beampatten_fig}      	       
	\end{subfigure}%
      \begin{subfigure}[b]{0.245\textwidth}
 		\centering       		
		\psfrag{Pfa}[Bl][Bl][0.6]{$P_\text{fa}$}
		\psfrag{Pd}[Bl][Bl][0.6]{$P_\text{d}$}
		\psfrag{phi  =  20}[Bl][Bl][0.35]{$\phi_\text{3dB}\!=\!20^\degree$}
		\psfrag{phi  =  25}[Bl][Bl][0.35]{$\phi_\text{3dB}\!=\!25^\degree$}
		\psfrag{phi  =  30}[Bl][Bl][0.35]{$\phi_\text{3dB}\!=\!30^\degree$}
		\includegraphics[width=42mm]{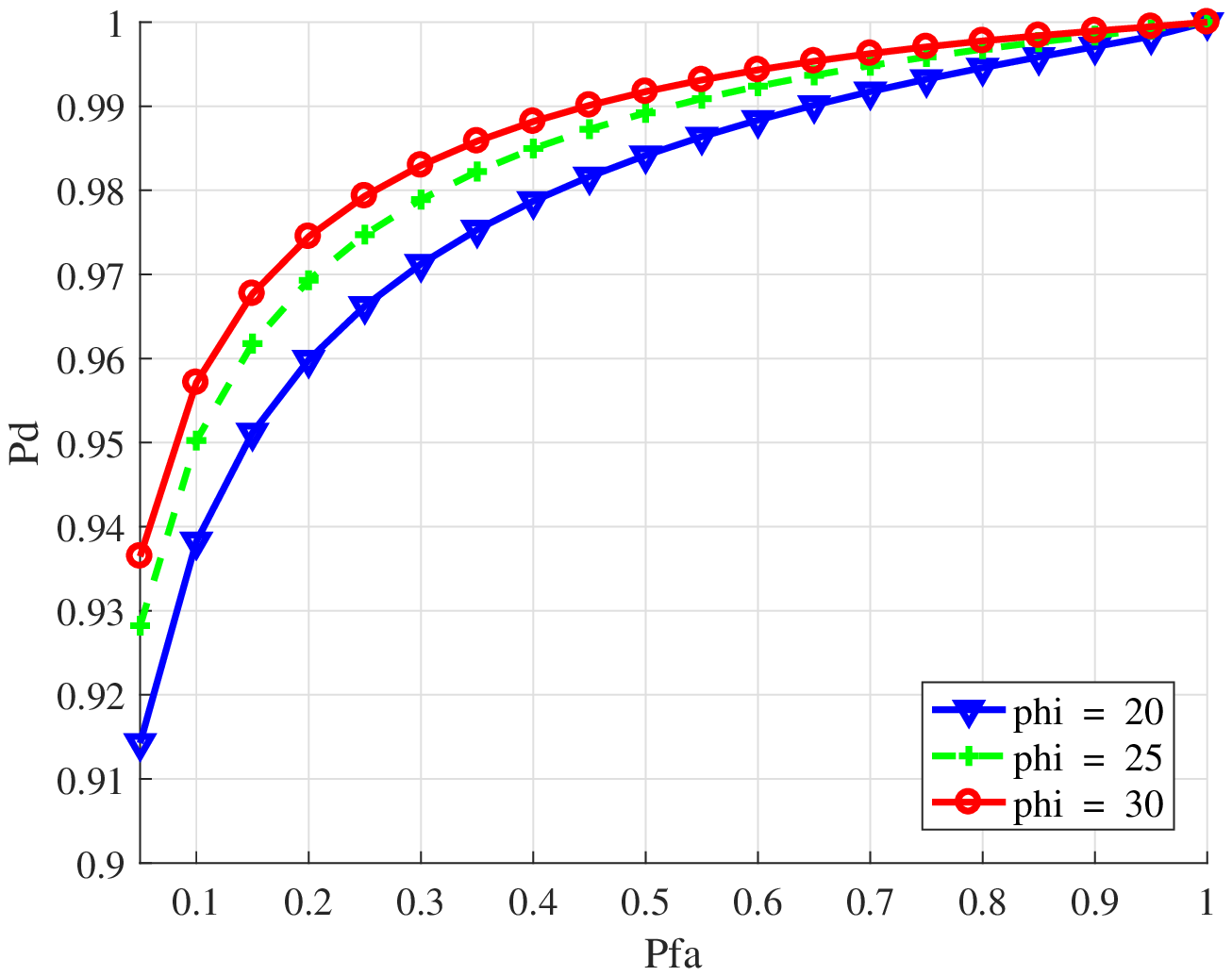}
		\caption{} 
		\label{Pd_Pf}      
      \end{subfigure} \\
\caption{(a) Beampattern in polar coordination, ~(b)  $P_\text{d}$ versus $P_\text{fa}$ for $\phi_\text{3dB}=20^\degree,25^\degree,30^\degree$.}
\vspace{2mm}
\end{figure}

 When beam-width  of sectors increases, RA spreads electromagnetic power in a wider area. To fairly compare the performance of our system with different $\phi_{3\text{dB}}$, we choose $A_0$ such that $E_A = 1$. The beampatterns of a sector of RA for $\phi_{3\text{dB}}=25^\degree, 35^\degree$ and the beampattern of a traditional omni-directional antenna are shown in Fig. \ref{beampatten_fig}. We can see that by decreasing $\phi_{3\text{dB}}$ and setting $E_A=1$, the gain of antenna in $0^\degree$ increases. Fig. \ref{Pd_Pf} shows the probability of detection versus the probability of false alarm for $N=16$, $M=8$ and $\phi_\text{3dB}=20^\degree,25^\degree,30^\degree$. We observe that as $\phi_\text{3dB}$ increases, 
$P_\text{d}$ increases. 
%
%
%
%
\par
\begin{figure}[!t]
\centering       		
\psfrag{phi  =  0}[Bl][Bl][0.6]{$\phi_\text{SR}\!=\!0^\degree$} 		
\psfrag{phi  =  10}[Bl][Bl][0.6]{$\phi_\text{SR}\!=\!10^\degree$} 	
\psfrag{phi  =  15}[Bl][Bl][0.6]{$\phi_\text{SR}\!=\!15^\degree$} 					
\psfrag{phi3dB}[Bl][Bl][0.7]{$\phi_\text{3dB}$}
\psfrag{Delta1}[Bl][Bl][0.7]{$\Delta_1$}
\includegraphics[width=65mm]{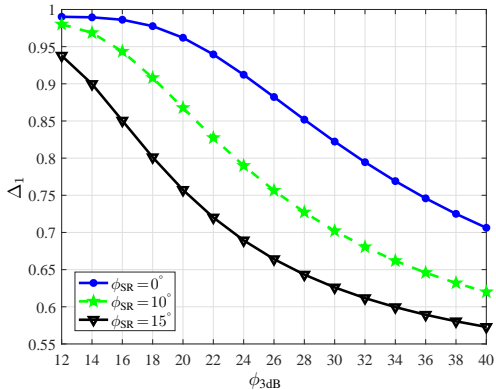}
\caption{$\Delta_1$ versus $\phi_\text{3dB}$.} 
\label{Delta1}      
\end{figure}

 The probability of first beam selection ($\Delta_1$) versus $\phi_\text{3dB}$ is plotted in Fig. \ref{Delta1} for $M=8$ when \SURx ~is located at $\phi_\text{SR}=0^\degree, 10^\degree, 15^\degree$. 
If we increase the beam-width $\phi_\text{3dB}$, $\Delta_1$ decreases, i.e., to increase $\Delta_1$, we need a narrower beam-width. 
Also, we note that when $\phi_\text{SR}$ approaches  $0^\degree$, the beam selection is more accurate.

\par
\begin{figure}[!t]
\vspace{-0mm}
\centering                             
\psfrag{phi  =  20}[Bl][Bl][0.6]{$\phi_\text{3dB}\!=\!20^\degree$}
\psfrag{phi  =  30}[Bl][Bl][0.6]{$\phi_\text{3dB}\!=\!30^\degree$}
\psfrag{Omni}[Bl][Bl][0.6]{Omni}
\psfrag{Capa}[Bl][Bl][0.7]{$\overline{C^\text{LB}_\text{Opt}}$}
\psfrag{Pav}[Bl][Bl][0.7]{$\Pbar$ [dB]}
\includegraphics[width=65mm]{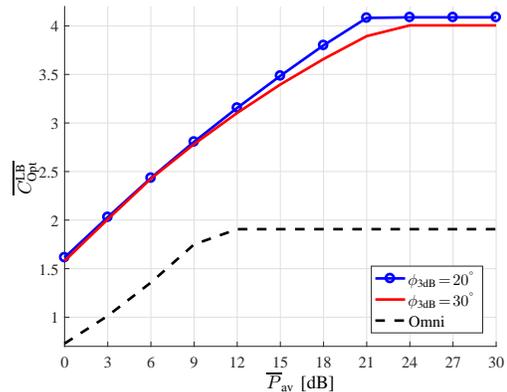}
\caption{Optimal Capacity averaged over orientation of \SURx ~and PU versus $\Pbar$.} 
\label{Capa_I0M8} 
\vspace{2mm}
\end{figure}

 Assume $\overline{C^\text{LB}_\text{Opt}}$ is the maximized capacity averaged over all possible orientations of \SURx ~($\phi_\text{SR}$) and PU ($\phi_\text{PU}$). Fig. \ref{Capa_I0M8} illustrates $\overline{C^\text{LB}_\text{Opt}}$ versus $\Pbar$ when $M=8, \Ibar=0$ dB, $\phi_\text{3dB} = 20^\degree, 30^\degree$. For comparison, the maximized  capacity when \SUTx ~equipped with omni-directional antenna and $\Phi$, $\zeta$ and $T_\text{sen}$ are optimized, 
is plotted. We can see that, in average, RA yields a higher capacity compared to omni-directional antenna. Also, a RA with smaller $\phi_\text{3dB}$ yields a higher  capacity, because it  can cancel more interference imposed on PU  from \SUTx. 
%
%
%
%
%
%
%
%
%
%
%
%
\par
\begin{figure}[!t]
\vspace{-0mm}
\centering
	\begin{subfigure}[b]{0.24\textwidth}                
		\centering
		\psfrag{phi  =  20}[Bl][Bl][0.4]{$\phi_\text{3dB}\!=\!20^\degree$}
		\psfrag{phi  =  30}[Bl][Bl][0.4]{$\phi_\text{3dB}\!=\!30^\degree$}		
		\psfrag{Pou}[Bl][Bl][0.55]{$\overline{P_\text{out}}$}
		\psfrag{Pav}[Bl][Bl][0.55]{$\Pbar$ [dB]}
		\includegraphics[width=42mm]{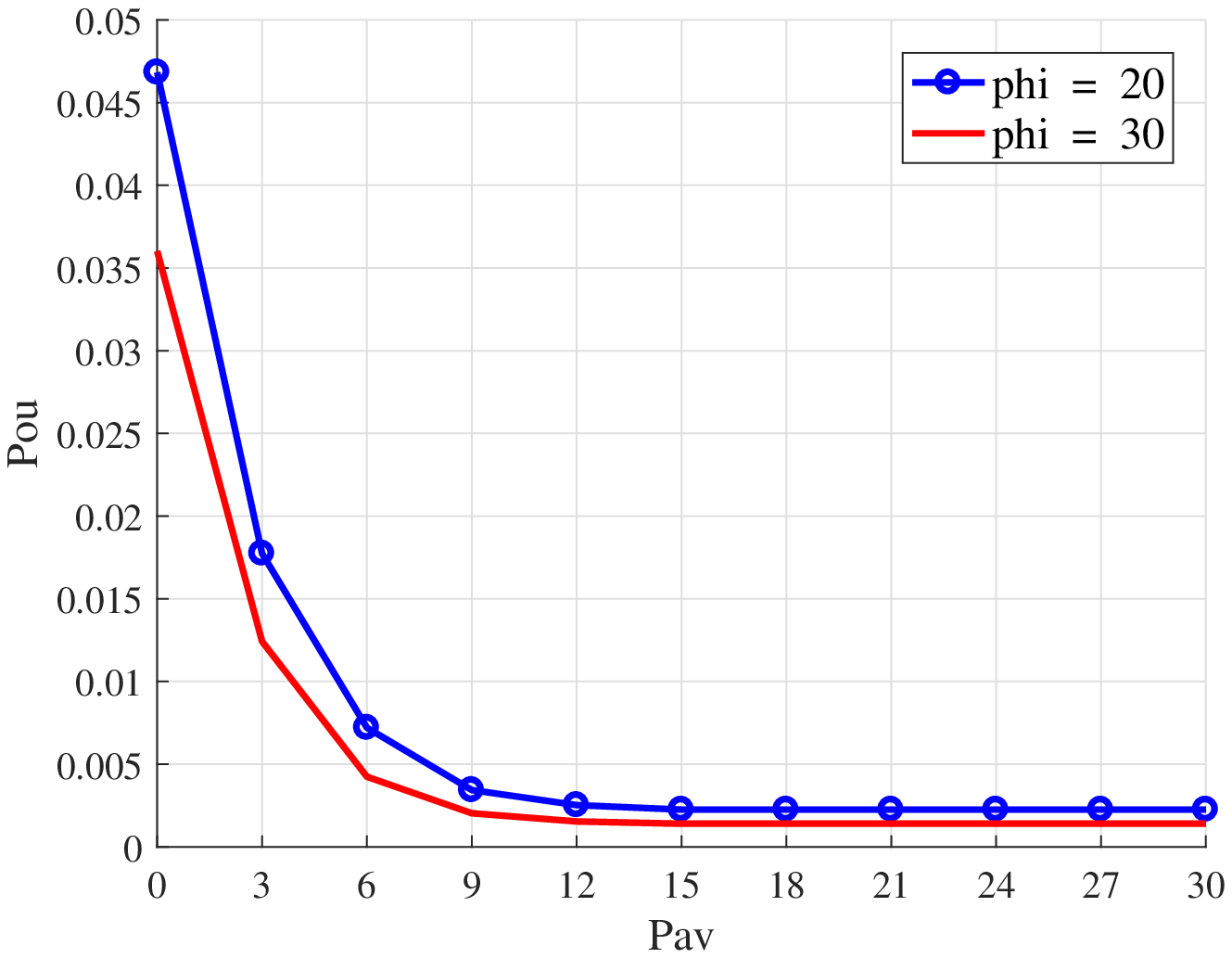}
		\caption{} 
		\label{Pout_Pav}       	       
	\end{subfigure}%
      \begin{subfigure}[b]{0.24\textwidth}
		\centering
		\psfrag{phi  =  20}[Bl][Bl][0.4]{$\phi_\text{3dB}\!=\!20^\degree$}
		\psfrag{phi  =  30}[Bl][Bl][0.4]{$\phi_\text{3dB}\!=\!30^\degree$}		
		\psfrag{Pe}[Bl][Bl][0.55]{$\overline{P_\text{e}}$}
		\psfrag{Pav}[Bl][Bl][0.55]{$\Pbar$ [dB]}
		\includegraphics[width=42mm]{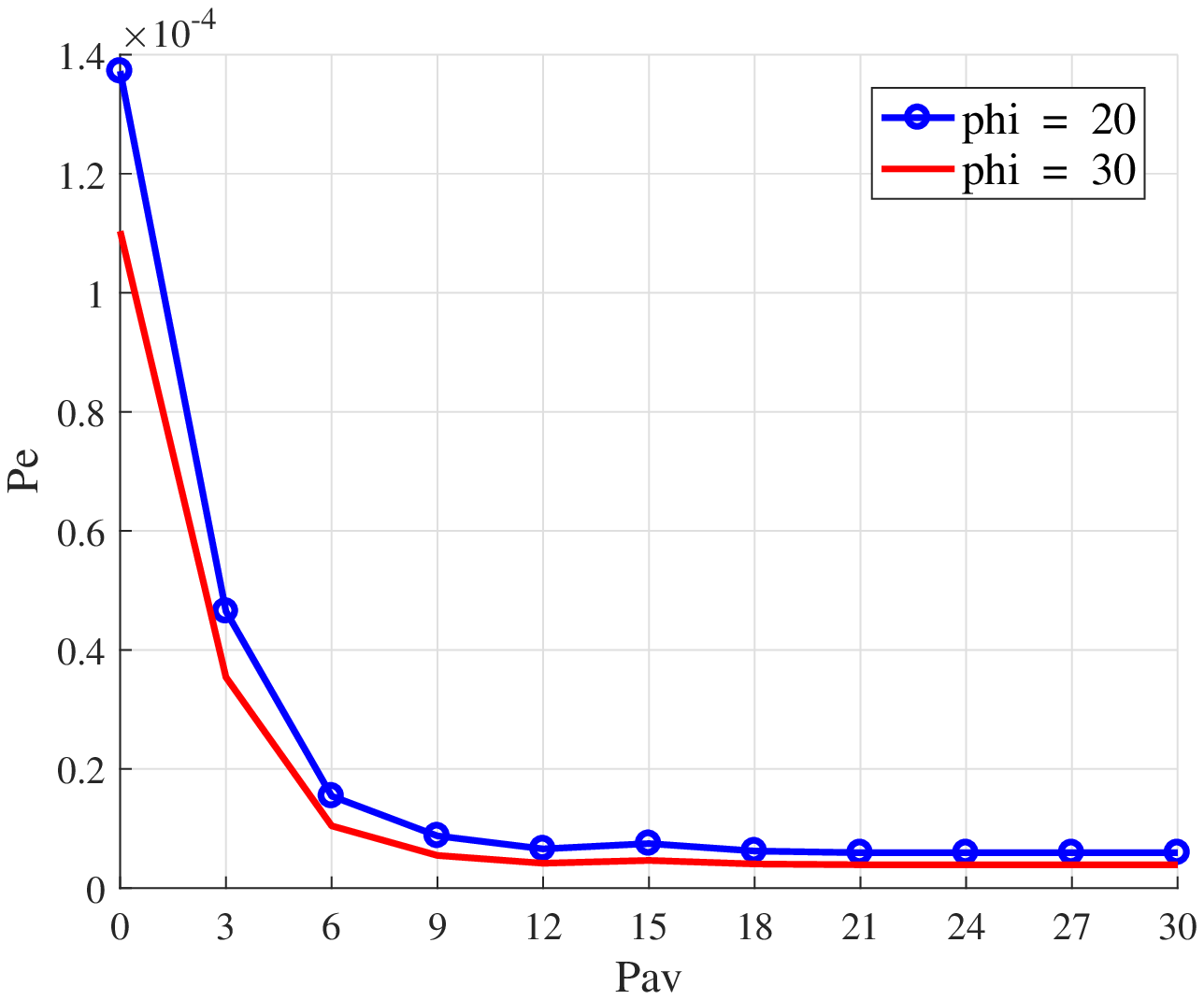}
		\caption{} 
		\label{Pe_Pav}      
      \end{subfigure} \\
\caption{ (a) $P_\text{out}$ versus $\Pbar$, ~(b) $P_\text{e}$ versus $\Pbar$.}
\vspace{0mm}
\end{figure}

 The averaged outage and symbol error probabilities over $\phi_\text{SR}$ and $\phi_\text{PU}$  (denoted as $\overline{P_\text{out}}$ and $\overline{P_\text{e}}$) are shown in Figs. \ref{Pout_Pav} and \ref{Pe_Pav}, respectively. Regarding the maximized capacity shown in Fig. \ref{Capa_I0M8}, we observe that increasing the beam-width from $20^\degree$ to $30^\degree$ yields lower outage and symbol error probabilities. However, the other performance metric, i.e. the ergodic capacity, decreases.
%
%
%
%
%
%
%
\section{Conclusion}

We considered a CR system consisting of a PU and a pair of \SUTx ~and \SURx.  The \SUTx ~is equipped with a RA which divides the angular space into $M$ sectors. The \SUTx ~first senses the activity of PU and transmits data to \SURx ~(if the channel is sensed idle) over the strongest  channel (sector) of the RA. We obtained the optimal channel sensing duration, the optimal power level and the optimal threshold, such that the ergodic capacity of CR system is maximized, subject to average interference and power constraints. In addition, we derived closed form expressions for outage and symbol error probabilities of our CR system.
%
\section*{Acknowledgment}
This research is supported by NSF under grant ECCS-1443942.
%
\bibliographystyle{IEEEtran}
\bibliography{ConfRef}
%
\end{document}